\documentstyle[12pt,aaspp4,flushrt]{article}
\newcommand{\sgr}{Sgr A$^*$}

\newcommand{\beq}{\begin{equation}}
\newcommand{\eeq}{\end{equation}}

\newcommand{\epm}{e^{\pm}}
\newcommand{\cpion}{\pi^{\pm}}

\newcommand{\NarrowMargins}{
  \setlength{\oddsidemargin}{+0.3in}
  \setlength{\evensidemargin}{-0.0in}
  \setlength{\textwidth}{6.2in}
  \setlength{\topmargin}{-0.75in}
  \setlength{\textheight}{9.25in}   }
\catcode`\@=11 
\def\lsim{\mathrel{\mathpalette\@versim<}}
\def\gsim{\mathrel{\mathpalette\@versim>}}
\def\@versim#1#2{\vcenter{\offinterlineskip
        \ialign{$\m@th#1\hfil##\hfil$\crcr#2\crcr\sim\crcr } }}
\catcode`\@=12 
\NarrowMargins
\begin{document}
\title{Probing the Two--Temperature Paradigm: \\
Observational Tests for the Basic Assumptions in ADAFs.}
\author{Rohan Mahadevan}
\affil{Institute of Astronomy, University of Cambridge, \\
 Madingley Road, Cambrdige, CB3 0HA, UK. \\
rohan@ast.cam.ac.uk}
\begin{abstract}
We calculate the flux and spectrum of synchrotron radiation produced by 
high energy electrons and positrons ($\epm$) in an advection dominated accretion flow 
(ADAF) around a black hole.  The $\epm$ are from the 
decay of charged pions which are created through proton--proton collisions.
We consider both a thermal and power--law energy distribution of 
protons, and show that the resulting $\epm$ synchrotron emission 
produces a characteristic spectrum between radio and X--ray frequencies. 
While previous signatures of the hot protons were only possible at gamma--ray
energies, via the production of gamma--rays through neutral pion decay, the present  results
provide a more observationally tractable way of probing the proton energy 
distribution and the two temperature structure in these accretion flows.
We discuss a number of strong observational predictions from these systems, 
as well as  the recent results of Mahadevan (1998) which appear to confirm 
the two temperature structure in ADAFs.  We show that the results provide 
support 
for both a power--law and thermal distribution of protons, with at least a third 
of the viscous energy going into the power--law.

\vspace{.2in}
\noindent {\bf Key words:} accretion, accretion discs -- elementary particles -- radiation mechanisms: nonthermal --
acceleration of particles -- Galaxy: centre  -- radio continuum: galaxies
\end{abstract}
\section{Introduction} 
In an advection--dominated accretion flow
(ADAF), all the viscous energy generated is stored in the accreting
gas and advected into the central star (Ichimaru 1977; Rees et
al. 1982; Abramowicz et al. 1998; Narayan \& Yi 1994,1995ab;
Abramowicz et al. 1995; see Narayan, Mahadevan \& Quataert 1998a for a
recent review).  If the central object happens to be a black hole, the
advected energy is lost through the event horizon, and the accreting
system appears very dim.  Most of the viscous heating is assumed to
mainly affect the ions, the more massive species, while the radiation
is produced primarily by the electrons.  Since the ions transfer only
small fraction of their energy to the electrons via Coulomb
collisions, the radiative efficiency of an ADAF is much less than the
total energy released during accretion (Rees et al. 1982).  The gas in
an ADAF forms a two temperature structure (Shapiro, Lightman, \&
Eardley 1976; Rees et al. 1982), with the ions attaining nearly virial
temperatures close to the black hole ($T_i \sim 10^{12}$K), and the
electrons being much cooler ($T_e \sim 10^{9.5}$K).

The success of ADAF models lies in their ability to accurately predict
the observed spectrum from a number of accreting stellar mass and
supermassive black hole systems (see Narayan et al. 1998a for a review).  
The spectrum depends
sensitively on the assumed energy distribution of the protons and
electrons as well as their individual cooling mechanisms.  The protons
cool by creating neutral pions, which subsequently decay into
gamma--rays (Mahadevan, Narayan, \& Krolik 1997), while the electrons
cool by various optically thin processes such as cyclo--synchrotron,
inverse Compton and bremsstrahlung radiation (see eg. Narayan \& Yi
1995b; Mahadevan 1997).  The electrons form a thermal distribution
(Mahadevan \& Quataert 1998), and give rise to a spectrum ranging from
radio to hard X--ray frequencies ($10^9$ Hz $ \lsim \nu \lsim 10^{21}$
Hz), but the energy distribution of the protons, and therefore the
resulting gamma--ray spectrum, is still unknown and depends on the
viscous heating mechanism involved (Mahadevan \& Quataert 1997;
Mahadevan et al. 1997).  If viscosity heats the protons into a thermal
distribution, the resulting gamma--ray spectrum is sharply peaked
around energies $\sim 70 $ MeV, while a power--law distribution
results in a power--law gamma--ray spectrum which extends to very high
energies (Mahadevan et al. 1997).

It is the purpose of this paper to probe the two temperature paradigm
further and attempt to resolve what the distribution function of the
protons is likely to be.  We are therefore interested in other
possible emission processes from the protons from which accurate
spectra can be determined. In particular, we focus on the production
of charged pions which subsequentially decay into electrons and
positrons.  These particles interact with the equipartition magnetic
fields in the accreting gas, and produce synchrotron radiation.
Although this energy leaves the plasma at lower frequencies than
gamma--rays, it provides the exciting possibility that current
sensitive and high resolution radio to X--ray telescopes might be used
to probe the much higher energy proton distribution function and
viscous heating mechanisms in these plasmas.

We begin in \S 2 by describing all the physical processes involved in
calculating the $\epm$ synchrotron spectra from both a thermal and
power--law parent proton distribution.  These calculations depend only
on the microphysics of particle collisions, decays, and emission
processes.\S 3 introduces the physical properties of an ADAF and,
using the results of \S 2, determines the general properties of the
resulting $\epm$ spectrum from an ADAF.
\S 4 applies these results to solar mass and supermassive 
accreting black holes.  In \S 5 we conclude by discussing the
implications of these results on the basic assumptions in ADAFs, and
propose future observational tests which would lead to a better
understanding of these systems.
\section{Physical Processes}
\subsection{Production of Electrons and Positrons}
The production of electrons and positrons through the decay of charged
pions is a three step process and has been studied by various authors
(see eg. Ginzburg \& Syrovatskii 1964; Ramaty \& Lingenfelter 1966ab,
1968; Stecker 1971; Bhadwar et al. 1977; Dermer 1986ab).  The
colliding protons produce a charged pion, $\cpion$, which decays into
a charged muon, $\mu^{\pm}$, and a muon neutrino, $\nu_{\mu}$.  The
charged muon decays into an electron, neutrino and their
antiparticles.  The reactions are
$$p + p \rightarrow \cpion + X, \ \ \ \ \cpion \rightarrow 
\mu^{\pm} + \nu_{\mu}(\bar{\nu}_{\mu}),$$
\beq
\mu^{\pm} \rightarrow e^{\pm} + \nu_e(\bar{\nu}_e) 
+ \bar{\nu}_{\mu}(\nu_{\mu}),
\eeq
where $X$ represents all other decay products.  The number of charged
pions created per unit volume, per unit time, with energy
$E_{\cpion}$, is given by (Landau \& Lifshitz 1975, \S 12;
cf. Mahadevan et al. 1997, eq.[4])
\begin{eqnarray}
R(E_{\cpion}) \equiv {dN_{\cpion} \over dE_{\cpion} \, dV} &=& {c
\over 2}
\, \int_1^{\gamma} \, d\gamma_1 \,
\int_{\gamma}^{\infty}  d\gamma_2 \,  \int_{-1}^{1} \, d\cos\!\theta  \ 
{d\sigma(E_{\cpion}; \gamma_1, \gamma_2, \cos\!\theta)\over
dE_{\cpion}} \,
\nonumber \\
&\times& n_{\gamma_1} \, n_{\gamma_2} \, \sqrt{ (\vec{\beta_1} -
\vec{\beta_2})^2 - (\vec{\beta_1} \times \vec{\beta_2})^2}
\ \ \mbox{s$^{-1}$},  
\label{cpionspectrum}
\end{eqnarray}
where ($\vec{\beta_1}, \vec{\beta_2}$), ($\gamma_1, \gamma_2$),
($n_{\gamma_1}, \ n_{\gamma_2} $) are the velocity parameters, Lorentz
factors, and densities of the two colliding particles respectively, 
$\cos\!\theta = \vec{\beta_1} \cdot \vec{\beta_2}/
|\vec{\beta_1}||\vec{\beta_2}|$,  $d\sigma(E_{\cpion}; \gamma_1,
\gamma_2, \cos\!\theta)/ dE_{\cpion}$ is the normalized differential
cross section for the production of a charged pion with energy
$E_{\cpion}$, and $c$ is the velocity of light.  Following Mahadevan
et al. (1997), the differential cross section is determined by using
the isobar model for energies $\lsim 4$ GeV (Lindenbaum \& Sternheimer
1957; Stecker 1971; Dermer 1986a), and the scaling model for energies
$\gsim 8$GeV (Badhwar et al. 1977; Stephens \& Bhadwar 1981; Dermer
1986b). At intermediate energies the differential cross section is
determined by interpolating smoothly (cubic interpolation) between the
two regimes.

The spectrum of electrons and positrons, $R(\epm)$, is derived from
the $\cpion$ spectrum by considering the reactions,
$$ \cpion \rightarrow \mu^{\pm} + \nu_{\mu}(\bar{\nu}_{\mu}), 
\ \ \ \mu^{\pm} \rightarrow
e^{\pm} + \nu_e(\bar{\nu}_e) + \bar{\nu}_{\mu}(\nu_{\mu}). $$
Ideally, determining the energy distribution of $\epm$ in the
laboratory frame requires three Lorentz transformations: one to rest
frame of the $\cpion$, to determine the decay spectrum of the
$\mu^{\pm}$, the second to the rest frame of the $\mu^{\pm}$, to
determine the decay spectrum of the $\epm$, and the third a
transformation back to the laboratory frame.  However, since the
Lorentz factor of the $\mu^{\pm}$ is small in the rest frame of the
$\cpion$,
\beq
\gamma_{\mu^{\pm}} =  {m_{\cpion}^2 + 
m_{\mu^{\pm}}^2 \over 2 m_{\cpion} m_{\mu^{\pm}}} \simeq 1.046,
\nonumber
\eeq
the difference between its velocity and that of the parent $\cpion$
can be neglected.  In this approximation, $R(E_{\cpion}) \simeq
R(E_{\mu^{\pm}})$, and determining the $\epm$ spectrum simplifies to
requiring only two Lorentz transformations.  $R(E_{\epm})$ can now be
obtained by performing the double integral (Ginzburg \& Syrovatskii
1964),
\beq
R(E_{\epm}) = \int \int {1 \over 2 \gamma_{\cpion}\beta_{\cpion} \,
p_{\epm}^{\prime}}
\, R(E_{\cpion}) \, {dP^{\prime} \over dE^{\prime}_{\epm}} \, dE_{\cpion} \, dE^{\prime}_{\epm},
\label{eprateeq}
\eeq
where the limits of integration are such that 
$$ \left | E_{\epm} - E_{\epm}^{\prime} \right | \leq
p_{\epm}^{\prime} \gamma_{\cpion} \beta_{\cpion}.$$

Here, $dP^{\prime} / dE^{\prime}_{\epm}$ is the probability of
creating an $\epm$ with energy $E^{\prime}_{\epm}$ and momentum
$p_{\epm}^{\prime}$ in the rest frame of the decaying muon, and is
given by (Ginzburg \& Syrovatskii 1964),
\beq
{dP^{\prime} \over dE^{\prime}_{\epm}} = {16 E_{\epm}^{\prime} \over
m_{\mu}^3 } \, p_{\epm}^{\prime} \, \left[ 3 - 4\, {E_{\epm}^{\prime}
\over m_{\mu}} \right].
\eeq
\subsection{Synchrotron Emission}
If the plasma in which the $\epm$'s are created has a magnetic field,
the particles interact with the magnetic field and cool by radiating
synchrotron emission.  The synchrotron radiation produces a
characteristic spectrum which depends directly on the energy
distribution of the $\epm$.  Since the distribution of $\epm$ is
determined by the proton energy distribution
(cf. eqs. [\ref{cpionspectrum}], [\ref{eprateeq}]), the resulting
synchrotron spectrum provides a probe to the energy distribution of
the protons.  Here, we provide the basic equations that determine the
$\epm$ synchrotron spectra, and show (\S2.3) how different initial
proton distributions give rise to different spectra.  In what follows
all quantities refer to the created $\epm$, unless otherwise noted.

Calculating the synchrotron emissivity, $L_{\nu}^s(\nu)$, requires
evaluating
\beq
L_{\nu}^s \, d\nu = \int^{\gamma_{\rm max}}_{\gamma_{\rm min}} \,
N(\gamma) \, j(\nu, \gamma) \, d\gamma \ \ \ \mbox{erg s$^{-1}$
cm$^{-3}$ Hz$^{-1}$},
\label{synchspectrum}
\eeq
where $\gamma$ is the Lorentz factor of the created $\epm$, $j(\nu,
\gamma)$ is the synchrotron emission formula averaged over all
particle directions, and $N(\gamma)$ is the steady state distribution
of $\epm$.  Since the $\epm$ created are highly relativistic (\S 2.3,
Fig. \ref{physical_fig}), and extend to very high energies, we can set
$\gamma_{\rm min} \gg 1$$,$ and  $\gamma_{\rm max} \rightarrow \infty$.
In this limit, $j(\nu, \gamma)$ takes the relativistic form
(eg. Rybicki \& Lightman 1979),
$$
j(\nu, \gamma) = S_0 F\left(\nu \over \nu_c \right), 
$$
\beq
S_0 = {4\pi \sqrt3 e^2 \nu_B \over 3\, c}, \ \ \ \nu_c = {3 \over 2}
\, \gamma^2 \nu_B,
\ \ \ \nu_B = {e B \over 2 \pi m_e c},
\label{synch}
\eeq
where we have averaged over all particle directions.  Here $F(x)$ is
an integral over modified Bessel functions (Rybicki \& Lightman 1979,
eq.[6.31c]), $S_0 = 1.56 \times 10^{-22} \, B $, and $\nu_B =
2.8\times 10^6$ Hz is the cyclotron frequency.  Integrating equation
(\ref{synch}) over all frequencies gives the total energy radiated per
particle,
\beq
\dot{E}_S  \equiv \int_0^{\infty} j(\nu, \gamma) \,
d\nu = 1.06 \times 10^{-15} \, \gamma^2 \, B^2 \ \ \
\mbox{erg s$^{-1}$}.
\label{totalsynch}
\eeq

For large $\gamma$, the $\epm$ radiate their energy very efficiently,
and their steady state distribution, $N(\gamma)$, is determined by the
competing effects of the ``creation'' and ``depletion'' of particles.
At a given energy $E$, the colliding protons produce $R(E)$ electrons
and positrons.  These $\epm$, however, cool very efficiently, thereby
reducing the number of $\epm$ at energy $E$.  The steady state
distribution can therefore be determined by solving the Fokker--Planck
equation without the stochastic term,
\beq
{\partial N(\gamma,t) \over \partial t} = - {\partial \over \partial
\gamma} \left[ \dot{\gamma} N(\gamma,t) \right] + R(\gamma,t)
\label{fokker},
\eeq
with $\partial N(\gamma,t)/ \partial t = 0$.  Here, $\dot{\gamma} =
\dot{E}_S/m_e c^2$ is the cooling rate of the $\epm$ and $R(\gamma)$
is their injection rate (cf. eq. [\ref{eprateeq}]).  In steady state,
the equation requires the net flux of particles between $\gamma_1$ and
$\gamma_2$ to be equal to the rate of injection of particles between
these two Lorentz factors.  Alternatively, the equation can be thought
of as a particle conservation equation in energy space with
$\dot{\gamma}$ being the velocity along the energy axis.

Setting $\partial N(\gamma,t)/ \partial t = 0$, equation
(\ref{fokker}) gives
$$
N(\gamma) = - {1 \over \dot{\gamma}} \, C(\gamma),
$$
\beq
C(\gamma) \equiv \int_\gamma^{\infty} R(\gamma^{\prime}) \,
d\gamma^{\prime},
\label{numbersynch}
\eeq
where $C(\gamma)$ is the total number of injected $\epm$ above Lorentz
factor $\gamma$.  Using equation (\ref{synchspectrum}), the resulting
synchrotron spectrum can be written as
\beq
L_{\nu}^s \, d\nu = \int^{\infty}_{\gamma_{\rm min}}  \, C(\gamma) \,
{ j(\nu, \gamma)  \over \dot{\gamma}} \, d\gamma  \ \ \ \mbox{erg cm$^{-3}$ s$^{-1}$ Hz$^{-1}$}.
\label{synchwithsteady}
\eeq
\subsection{Synchrotron Spectra from Electrons and Positrons.}
Using equations (\ref{cpionspectrum}), (\ref{eprateeq}),
(\ref{numbersynch}), and (\ref{synchwithsteady}), the synchrotron
spectra from the $\epm$ can be determined, and depends only on the
initial proton energy distribution.  
We consider here two
extreme distributions: a relativistic Maxwell--Boltzmann, $N(E)
\propto E^2 \exp(-E/kT)$, and a power--law with energy index $s$,
$N(E) \propto E^{-s}$.  In almost all areas of astrophysics, the true
proton distribution is probably some combination of the two.

Before determining the exact synchrotron spectra, we point out two
general properties of equation (\ref{synchwithsteady}).  First, the
total synchrotron power, $P_{\rm total}$, is given by
\beq
P_{\rm total} = \int_0^{\infty} L_{\nu}^s \, d\nu =
m_e c^2 \, \int^{\infty}_{\gamma_{\rm min}}  \, C(\gamma) \, d\gamma \ \ \
\mbox{erg s$^{-1}$},
\label{totalsynch2}
\eeq
which is equal to the total energy of $\epm$ that are created per
second.  This depends only on the injected energy spectrum and is
independent of the synchrotron emissivity, $j(\nu,\gamma)$, and
therefore of the magnetic field.  We might expect that since particles
in a very low magnetic field radiate synchrotron emission
inefficiently (cf. eq.[\ref{totalsynch}]), the total synchrotron
emission from this plasma would be low.  However, since we are
interested in steady state distributions, the decrease in the amount
of cooling is exactly compensated by an increase in the steady state
number density (cf. eq. [\ref{numbersynch}]).  Similarly, higher
magnetic fields lead to more efficient cooling, which results in a
lower steady state number density of particles.  The two effects
exactly compensate for each other and leaves the total cooling rate
unchanged.

Second, the synchrotron spectrum, in a $\nu L_{\nu}$ versus $\nu$
plot, is only a function of the dimensionless frequency $\chi \equiv
\nu / \nu_B$ (cf. eqs. [\ref{synch}, \ref{synchwithsteady}]).  
This shows that while the intrinsic shape of the synchrotron spectrum
does not change as the magnetic field changes, the position of the
spectrum does move horizontally to the right (left) as the magnetic
field is increased (decreased).
\subsubsection{Thermal Proton Distribution}
Using a normalized relativistic Maxwell--Boltzmann distribution for
$n_{\gamma_1}, n_{\gamma_2}$ in equation (\ref{cpionspectrum}),
$R(E_{\cpion})$ is obtained by performing the integrals given in
Dermer (1986a; see also Mahadevan et al. 1997, eq. [12] with the
appropriate changes for $\cpion$ cross--sections).  The results are
then used in equation (\ref{eprateeq}) to determine the resulting
production spectrum, $R(E_{\epm})$, of electrons and positrons.

Figure \ref{physical_fig}a shows $R(E_{\epm})$ for different values of
the dimensionless proton temperature $\theta_p = kT_p/m_p c^2 = 0.05,
0.1, 0.2$.  The spectra rise at low energies, peak near $E_{\epm} \sim 33$~MeV, 
and then decline exponentially at higher energies.

An important feature shown in Figure \ref{physical_fig}a is the effect
of small changes in temperature on $R(E_{\epm})$.  At low
temperatures, only protons in the exponential tail of the
Maxwell--Boltzmann distribution have energies above threshold to
produce pions and therefore electrons and positrons.  Increasing the
temperature by a factor of two from 0.05 to 0.1, dramatically
increases the number of protons with energies above threshold, thereby
increasing $R(E_{\epm})$ by more than an order of magnitude. Changing
the temperature again, from 0.1 to 0.2, has a less drastic effect on
the production rate.  At these high temperatures the protons
responsible for most of the $\cpion$ production are no longer in the
tail of the Maxwell--Boltzmann distribution, but rather have energies
of order the average energy of the plasma.  Therefore changing the
temperature by a factor of two, has the effect of increasing
$R(E_{\epm})$ by nearly the same amount.

Figure \ref{physical_fig}b shows the resulting steady state
distribution of $\epm$ (cf. eq.[\ref{numbersynch}]). At low energies,
the product $\dot{\gamma} N(\gamma)$ is a constant, which gives a
power--law distribution $N(\gamma) \propto \gamma^{-2}$, while at
higher energies $N(\gamma)$ decreases exponentially.  The low energy
power--law is the result of high energy $\epm$ that lose their energy
due to efficient synchroton cooling.

Figure \ref{physical_fig}c shows the resulting $\epm$ synchrotron
spectrum for three values of the magnetic field $B = 10, \, 10^4, \,
10^8$ Gauss (solid, dashed, and dotted lines respectively), and for
$\theta_p = 0.2$ (lower values of $\theta_p$ result in identical
spectra, but with much lower fluxes).  As expected, increasing the
magnetic field merely shifts the spectrum to higher frequencies, and
leaves the total flux and shape of the spectrum unchanged. The change
in the steady state distribution $N(\gamma)$ from a power--law at low
energies ($\propto \gamma^{-2}$) to an exponential at high energies,
results in a synchrotron spectrum that rises as $\nu L_{\nu} \propto
\nu^{+0.5}$, and then turns over exponentially (Rybicki \& Lightman
1979).  The frequency, $\nu_t$, at which the spectrum turns over is
determined by the Lorentz factor of the $\epm$ responsible for most of
the radiation ($\gamma \sim 100$, see Fig. \ref{physical_fig}b), as
well as the magnetic field strength, and is given by $\nu_t \sim 3
\times 10^6 \, \gamma^2 \, B \sim 10^{10} \, B $.
\subsubsection{Power--law Proton Distribution}
Using a normalized power--law proton distribution, $n_{\gamma} = (s-1)
\gamma^{-s} $, with energy index $s$, equations (\ref{cpionspectrum})
and (\ref{eprateeq}) are used to determine the production rate,
$R(E_{\epm})$, of electrons and positrons.  The calculation is
identical to that of $\pi^0$ production with the appropriate changes
for $\cpion$ cross--sections (see Mahadevan et al. 1997).

Figure \ref{physical_fig}d shows $R(E_{\epm})$ for three values of the
energy index $s$.  The spectrum rises at low energies, turns over at
$E_{\epm}\sim 35$ MeV, and extends as a power--law, $E^{-s}$, with the
same energy dependence as the parent proton distribution (Ginzburg \&
Syrovatskii 1964).  While the energy at which the spectra turn over
are similar to the thermal case (cf. Fig. \ref{physical_fig}a), the
non--thermal spectra are much broader and extend to very high
energies.  In addition, the production rate is not very sensitive to
the energy index $s$.  An increase in $s$ results in a small decrease
in the number of protons above the threshold energy, which moderately
decreases the total production rate.

Figure \ref{physical_fig}e shows the resulting steady state $\epm$
distribution.  Similar to the thermal case, the product $\dot{\gamma}
N(\gamma)$ is constant at low energies which gives a power--law
distribution $N(\gamma) \propto \gamma^{-2}$.  At higher energies,
however, the shape of the distribution deviates substantially from the
thermal case, and extends to very high energies as $N(\gamma) \propto
\gamma^{-(s-1)}$ (cf.  eq. [\ref{numbersynch}]).

The resulting $\epm$ synchrotron spectrum is shown in Figure
\ref{physical_fig}f for three values of the magnetic field $B = 10, \,
10^4, \, 10^8$ Gauss, and for $s = 2.75$.  As in the thermal case,
increasing the magnetic field merely shifts the spectrum to higher
frequencies and leaves the total flux and shape of the spectrum
unchanged.  The change in the index of the steady state distribution
from low to high energies, results in a synchrotron spectrum that
rises as $\nu L_{\nu} \propto \nu^{+0.5}$, which then steepens to $\nu
L_{\nu} \propto \nu^{-s/2}$ (Rybicki \& Lightman 1979).  Therefore,
while different values for the proton energy index $s$ result in
identical spectra at low frequencies, smaller values of $s$ give rise
to harder spectra at high frequencies.  For all $s$, the frequency at
which the spectrum turns over is given by, $\nu_t \sim 3 \times 10^6
\, \gamma^2 \, B \sim 10^{11} \, B $, where we have set $\gamma \sim
200$ (cf. Fig. \ref{physical_fig}d).
\section{Application to ADAFs} 
The results from the previous section
can now be used to determine the total $\epm$ synchrotron spectrum
from an ADAF.  Since an ADAF is well approximated by a series of
concentric spherical shells (Narayan \& Yi 1995ab), with the properties
of the gas varying with radius, the total $\epm$ spectrum is obtained
by determining the synchrotron spectra from each shell, and then
propagating the resulting spectra through the accretion flow to the
observer (\S 3.2).  Determining the spectrum from each shell requires
a knowledge of the proton densities and magnetic field strengths at
each radius.  These quantities are obtained by solving for the global
structure of an ADAF.
\subsection{ADAF Equations and Properties}
The global structure of an ADAF is characterized by four parameters:
the viscosity parameter $\alpha$ (Shakura \& Sunyaev 1973), the ratio
of gas to total pressure $\beta_{\rm adv}$, the mass of the central
black hole $M$, and the accretion rate $\dot{M}$.  Given these
parameters, the global structure of an ADAF can be calculated ( Gammie \& Popham 1998; 
Popham \& Gammie 1998; Narayan, Kato, \& Honma 1997; Chen, Abramowicz, \&
Lasota 1997), and for the results presented here, we use the new fully
relativistic, self--consistent, global solutions in the Kerr metric
developed by Popham \& Gammie (1998).  In general, ADAFs appear to
have large values of $\alpha \gsim 0.1$, and the uncertainty in the
value is a factor of a few (Narayan 1996).  Most ADAF models in the
literature use $\alpha = 0.3$ (see Narayan et al. 1998a).  For the
magnetic field strength, we consider two extreme values of $\beta_{\rm
adv} = (0.5, 0.9)$, which corresponds, respectively, to an
equipartition field, and to one where the field contributes negligibly
($\sim 1/10$) to the total pressure.

Assuming that the mass fraction of
hydrogen is $X = 0.75$, the total numberdensity of protons is obtained from the
continuity equation,\footnote{The continuity
equation is given by,
$$ \dot{M} = 4\pi R^2 \rho(R) v(R) \, \cos\theta_{\rm H} \, 
\left( {1-R_s/R \over 1-[v(R)/c]^2} \right)^{1/2}.$$ }
and is given by,
\begin{eqnarray}
n_p(r) &=& {X\rho(r) \over m_p} = 1.9 \times 10^{19} m^{-1} \dot{m}
r^{-2}
\left[{v(r) \over c} \right]^{-1} \, {1\over \cos\theta_{\rm H}} \,
\left( 1 - [v(r)/c]^2 \over 1 - r^{-1} \right)^{1/2} \nonumber \\
&\equiv& m^{-1} \dot{m} \tilde{n}(r) \ \ \mbox{cm $^{-3}$,}
\label{ntilde}
\end{eqnarray}
where $\rho(r)$ is the mass density and $v(r)$ is the radial velocity
of the gas.  Here, $m = M/M_{\odot}$ is the mass of the black hole in
solar mass units, $\dot{m} = \dot{M}/\dot{M}_{\rm Edd}$ is the
accretion rate in Eddington units with with $\dot{M}_{\rm Edd} = 10 \,
L_{\rm Edd}/c^2 = 1.38 \times 10^{18} m $ g s$^{-1}$, and $r = R/R_s$
is the radius in Schwarzschild units where $R_s = 2GM/c^2 = 2.95
\times 10^5 \, m$ cm.  This equation differs from that in Mahadevan et
al. (1997) by the two last terms on the right.
The factor $(\cos\theta_{\rm H})^{-1}$ accounts for the slight
non--spherical geometry of the flow (Narayan, Barret \& McClintock
1997a, Appendix A), and the term in parentheses is a general
relativistic correction factor (Popham \& Gammie 1998).  We obtain
$v(r)/c$ and $\cos\theta_{\rm H}$ from the global ADAF solutions of
Popham \& Gammie (1998).

The magnetic field strength is determined by the parameter $\beta_{\rm
adv}$, and is defined by (Narayan \& Yi 1995b),
$$ {B^2(r) \over 24\,\pi} = \left( {1 - \beta_{\rm adv} \over \beta_{\rm adv}}
\right) \, p_g,$$
\beq
p_g = \beta_{\rm{adv}} \rho(r) \, c^2_s(r) = {\rho(r) k \, T_p \over
                \mu_i \, m_p} + {\rho(r) k \, T_e \over \mu_e \, m_p},
\label{tempeq}
\eeq
where $p_g$ is the gas pressure, $c_s(r)$ is the isothermal sound
speed, and $\mu_i = 1.23, \ \mu_e = 1.14$ are the effective molecular
weights of the ions and electrons respectively.  Since $T_p \gg T_e$
we neglect the second term to obtain
\beq
T_p(r) = 1.34 \times 10^{13} \, \beta_{\rm{adv}} \, \left[ {c_s(r)
                \over c} \right]^2 \ \ \mbox{K}. \label{teeq}
\eeq
We obtain  $c_s(r)/c$  from the global ADAF solutions (Popham \& Gammie 1998).

Given $n_p(r)$ and $B(r)$ from equations (\ref{ntilde}), (\ref{tempeq}), and (\ref{teeq}),
the $\epm$ spectrum can now be determined. 
\subsection{ADAF Spectra}
In addition to the numberdensity of protons and magnetic field 
strength, the spectrum from each radius, and therefore the total $\epm$
synchrotron spectrum from the ADAF, also depends on the energy
distribution of the protons (\S 2.3).  Since our knowledge of the
viscous heating of protons is unknown, it is still uncertain as to
whether the process leads to a thermal or power--law distribution in
proton energies, or possibly a combination of the two.  Below, we
explore both distributions and calculate the $\epm$ synchrotron
spectrum associated with them.

At each radius, we write the numberdensity of protons as a function of
energy, $n_p(r,\gamma)$, as the product of a normalized velocity
distribution, which depends on the temperature at each radius, and the
total numberdensity of protons (see eg. Mahadevan et al. 1997),
\beq
n_p(r,\gamma) = n_p(r) \, n_{\gamma}[\gamma, \theta_p(r)], \label{3dreq8}
\eeq
with
\beq
\int_{1}^{\infty}  n_{\gamma}[\gamma, \theta_p(r)] \, d\gamma = 1, \ \ \ \ \
\int_1^{\infty} (\gamma - 1) \, n_{\gamma}[\gamma, \theta_p(r)] \, d\gamma =  {3 \over 2} \,
\theta_p(r).
\label{fdeq10}
\eeq

The total numberdensity of protons, $n_p(r)$, is determined from
equation (\ref{ntilde}), and for the present results we consider a
normalized velocity distribution, $n_{\gamma}[\gamma, \theta_p(r)]$
which is either a relativistic Maxwell--Boltzmann or power--law
distribution.  With this functional form of $n_p(r,\gamma)$, equations
(\ref{cpionspectrum}) and (\ref{eprateeq}) can be evaluated to give
the total production of $\epm$, and therefore the total synchrotron
emission.

Determining the synchrotron spectrum from the $\epm$ in an ADAF,
requires three additional pieces of physics to those presented in \S 2.
First, the emergent spectrum at radii close to the black hole must be
corrected for gravitational redshift effects. This is included in the
calculations that follow, and has the effect that $L_{\nu}^s(\nu)$
observed at large radii is now $L_{\nu}^s[\nu(1-1/r^{1/2})]$.  This
shifts $L_{\nu}^s$ redward by the gravitational redshift factor.  The
inclusion of gravitational redshift will change the entire optically
thin synchrotron spectrum determined in
\S 2.3.  However, this will only affect the spectrum dramatically within a 
few Schwarzschild radii.

Second, in addition to calculating the optically thin synchrotron
spectrum at each radius in the flow (\S 2.3), absorption processes
also need to be included.  This will change the resulting synchrotron
spectrum at frequencies where the plasma is optically thick. In
particular, since the thermal electrons in an ADAF radiate highly
self--absorbed synchrotron emission (Narayan \& Yi 1995b; see also
Mahadevan 1997), these electrons will absorb any radiation that
is above the local black body value.  Therefore, to account for this
absorption process we replace any ``excess'' synchrotron emission from
the $\epm$ with the local black body spectrum at each radius.  In a
highly self--absorbed plasma, this is a good approximation to the
exact radiative transfer calculations.

In addition to the absorption process described above,
self--absorption by the created $\epm$ is also included.  However, in
the calculations that follow, we find that absorption from the
thermal electrons is always more important.  Since the amount of
absorption depends on the physical size of the emitting region, the
calculations for the rest of this section are for a black hole mass $m
= 10^{6}$, accreting at a rate $\dot{m}=10^{-4}$.
\S 4 explores the results for different black hole masses and accretion rates.

Finally, the calculations and spectra discussed in \S 2 assumes that a steady
state distribution has been established. The only timescale of 
interest there was the synchrotron  cooling time.  However, 
in an ADAF, the accretion time, $t_{ac}$, is another 
important timescale in the problem.
If the synchrotron cooling time is longer than the accretion time, a steady
state distribution  will not be able to form, and while the $\epm$ will 
still radiate synchrotron emission, the spectral results of \S 2
are no longer valid.  Setting the synchrotron cooling time $t_s =
\gamma/\dot{\gamma}_s \simeq 7.4 \times 10^{8} / \gamma \, B^2$, equal
to the accretion time $t_{ac} \simeq 1.8\times 10^{-5} \, \alpha^{-1}
\, m r^{3/2}$ (Mahadevan \& Quataert 1998), gives
\beq
\dot{m} \gsim 4 \times 10^{-4} \,
\left( {\alpha \over 0.3} \right)^2 \,
\left({\gamma \over 200} \right)^{-1} \,
\left({ r \over 10^4} \right) \, \left( {1-\beta_{\rm adv} \over 0.5 }
\right)^{-1},
\eeq
where we have used the self--similar scaling of the magnetic field
with radius (Narayan \& Yi 1995b; see also Mahadevan 1997), and have
set $\gamma \sim 200$, the Lorentz factor of the $\epm$ responsible 
for most of the radiation (cf. Fig. \ref{physical_fig}e) .  For
systems where $\dot{m} \gsim 10^{-4}$, the timescale for the $\epm$ 
to cool via synchrotron radiation is less than the accretion time, and a steady
state distribution will be established.  The results of 
\S 2 are therefore valid.  For lower accretion rates,
the spectrum will be modified, but we do not consider this regime since 
the luminosities from these systems will be too low to be of observational consequence.
All published models of ADAFs in the literature have $\dot{m} > 10^{-4}$.
\subsubsection{Thermal Distribution}
Using equation (\ref{ntilde}) the synchrotron emissivity per unit
scaled volume in an ADAF is given by
\beq
L_{\nu} = {d E \over dt \,  d\tilde{V} \, d\nu} =
2.57 \times 10^{16} \, m \dot{m}^2 \,  \tilde{n}_p^2(r) \, L_{\nu}^s \ \ \ 
\mbox{erg s$^{-1}$ Hz$^{-1},$}
\label{thermalpion}
\eeq
where $L_{\nu}^s$ is the synchrotron emissivity for $n_p = 1$
cm$^{-3}$ (cf. eq. [\ref{synchwithsteady}]), and $dV = (2.95\times
10^5)^3 \, m^3 \, d\tilde{V} $. Here, $d\tilde{V} = 4\pi r^2 \, dr$,
is the scaled volume in Schwarzschild units.  The dependence of
$L_{\nu}$ on the energy distribution of the protons is taken into
account by the form of $R(E_{\epm})$ that is used to evaluate
$L_{\nu}^s$ (cf. Figure \ref{physical_fig}a).  As expected, equation
(\ref{thermalpion}) shows that the luminosity of the optically thin
part of the spectrum increases as $m \dot{m}^2$ similar to the
$\gamma$--ray spectrum (Mahadevan et al. 1997).  However, this scaling
is no longer valid for frequencies where the spectrum is
self--absorbed.

All the emission is produced in the inner most regions of the flow ($r \lsim
5$) where the proton temperatures are high enough to produce pions.
At larger radii, the temperatures drop rapidly, and almost none of the
protons have energies above the threshold for pion production (cf. \S
2).  Since the $\epm$ synchrotron emission originates from a few
Schwarzschild radii, all the self--absorption is from the thermal
electrons at $r\lsim 5$.  

Figure 2a shows the results for two values of the viscosity parameter
$\alpha$ (0.1 and 0.3) and two values of $\beta_{\rm adv}$ (0.5 and
0.9).  At relatively high frequencies, the spectra are optically
thin and show the characteristic rise ($\nu L_{\nu} \propto \nu^{+0.5}$) 
and exponential turnover expected from a thermal 
proton distribution (\S 2, Fig. \ref{physical_fig}c).  At lower frequencies,  however,
the emission is highly self--absorbed by the ambient thermal electrons.
The resultant self--absorbed spectrum is
Rayleigh--Jeans and has a spectral dependence $\nu L_{\nu} \propto
\nu^{3}$.  Although the distinct spectral change offers an interesting
testable prediction, the luminosity from this process is negligible
compared with the synchrotron luminosity from the thermal electrons at
these frequencies.  The thermal $\epm$ spectrum is therefore
undetectable (see \S 4).

For a given $\alpha$, increasing $\beta_{\rm adv}$ leads to an
increase in gas pressure, which therefore increases the proton
temperature (cf. eq.[\ref{teeq}]). Changing $\beta_{\rm adv}$ from 0.5
to 0.9, causes the temperature to increase by a factor $\sim 2$, and
since the $\epm$ flux is extremely sensitive to temperature (cf. \S
2), the synchrotron flux increases by nearly 2 orders of magnitude.
For a fixed $\beta_{\rm adv}$, changing $\alpha$ also affects the synchrotron luminosity.  Since
$\cpion$ production scales as $n_p^2 \propto \alpha^{-2}$ (Narayan \&
Yi 1995b), we might expect that decreasing $\alpha$ increases the
total luminosity.  In the case of high $\beta_{\rm adv}$ this is the case, 
but in the more detailed global solutions in the Kerr metric, the
slight change in temperature with $\alpha$ in the innermost regions of
the flow, for low $\beta_{\rm adv}$,  compensates for the increase in numberdensity, 
and the net result is a decrease in the total flux.
\subsubsection{Power--Law Distribution}
For a power--law distribution of protons, the total number of protons
is still given by equation (\ref{ntilde}), but the distribution
$n_{\gamma}[\gamma, \theta_p(r)]$ is no longer Maxwellian.  Following
Mahadevan et al. (1997) we write the energy distribution of protons as
\beq
n[\gamma, \theta_p(r)] \, d\gamma = \left\{ [1-\zeta(r)] \,
\delta(\gamma - 1) + (s-1) \, \zeta(r) \, \gamma^{-s}\right\} \,
d\gamma, \label{ntdef}
\eeq
where a fraction $1 - \zeta(r)$ of the protons have $\gamma \sim 1$,
and a fraction $\zeta(r)$ are in a power--law tail with index $s$.
The fraction $\zeta(r)$ is fixed by the energy requirement
(cf. eq. [\ref{fdeq10}]):
\beq
\zeta(r)  = {3 \over 2 } \, (s-2) \,  \theta_p(r). \label{zetaeq}
\eeq
Using equation (\ref{ntdef}) in equation (\ref{cpionspectrum}) gives
the total charged pion spectrum.  In an ADAF, $\zeta \ll 1$, and
Mahadevan et al. (1997) have shown that most of the contribution to
the pion spectrum comes from protons in the power--law tail colliding
with protons at rest.  Equation (\ref{ntdef}) can therefore be
simplified to
$$
n[\gamma, \theta_p(r)] \, d\gamma \simeq (s-1) \, \zeta(r) \, \gamma^{-s} \, d\gamma,
$$
and the $\epm$ synchrotron emissivity per unit scaled volume takes the
form
\beq
L_{\nu} = {d E \over dt \,  d\tilde{V} \, d\nu} =
2.57 \times 10^{16} \, m \dot{m}^2 \,  \tilde{n}_p^2(r) \,\zeta(r) \, [1-\zeta(r)] \,  
L_{\nu}^s,
\label{nonthermalpion}
\eeq
where $L_{\nu}^s$ represents the synchrotron emissivity using a proton
distribution $n(\gamma) = (s-1) \gamma^{-s}$
(cf. eq.[\ref{synchwithsteady}] and Fig. \ref{physical_fig}f).

Figure 2b shows the resulting spectrum for a proton energy distribution
with power--law index $s=2.75$ for different values of $\alpha  = (0.1, 0.3)$ 
and $\beta_{\rm adv}= (0.5,0.9)$. Similar to the thermal case, the optically
thin part of the spectrum scales as $m \, \dot{m}^2$
(cf. eq.[\ref{nonthermalpion}]), but at lower frequencies this scaling
is no longer valid since the spectrum is self--absorbed.   For a fixed 
$\alpha$ increasing $\beta_{\rm adv}$ decreases the magnetic field which 
shifts the spectrum to lower frequencies (\S 2.3, Fig. \ref{physical_fig}f).
For fixed  $\beta_{\rm adv}$ decreasing $\alpha$ increases the numberdensity 
($n_p \propto \alpha^{-1}$) which increases the optically thin flux.

Unlike the thermal case, the emission comes from nearly all radii in the ADAF.
In particular, since the protons are assumed to be accelerated into a
power--law at all radii, most of the high frequency emission
originates from $r \lsim 30$ while the low frequency emission comes
from $r\lsim 10^4$.  To understand this effect, recall that an $\epm$
with Lorentz factor $\gamma$ radiates most of its energy at
frequencies $\nu \propto \gamma^2 \, B$.  Since $\gamma$ is fixed at
all radii by the microphysics of particle decays (\S 2 and
Fig. \ref{physical_fig}d), the frequency at which most of the
synchrotron radiation emerges depends only on the magnetic field.  As
the magnetic field decreases with increasing radius, lower frequency
emission occurs farther away from the black hole.

At relatively high frequencies, the spectra are optically
thin and show the characteristic rise ($\nu L_{\nu} \propto \nu^{+0.5}$), turnover, 
and power--law tail ($\nu L_{\nu} \propto \nu^{-s/2}$)
expected from a power--law proton distribution (\S 2, Fig. \ref{physical_fig}f).  
At lower frequencies,  however,
the emission is highly self--absorbed by the ambient thermal electrons, and 
predicted spectrum is determined by self--absorption at all radii (\S 3.2).
We show below (\S\S 4 and 5) that the expected spectral shape provides 
an interesting prediction for the ADAF paradigm, and in the 
case of our Galactic Centre, Mahadevan (1998) has shown that these predictions
do indeed agree quite well with the observations. 
\section{Results: Application to Solar and Supermassive Black Holes}
This section combines the spectra produced by both the protons and 
electrons in an ADAF. 
The bulk of the electrons cool via thermal synchrotron,
bremsstrahlung and inverse Comptonization (Narayan \& Yi 1995b;
see also Mahadevan 1997), while the protons cool via the
production of neutral and charged pions.  In the results that 
follow, the $\gamma$--ray spectrum, 
produced by the decay of neutral pions, 
has been updated from Mahadevan
et al. (1997) to include general relativistic effects, gravitational
redshift, and the non--spherical geometry of the flow (\S 3).
\subsection{Thermal Distribution}
For a thermal distribution of protons, 
Figure \ref{thermalfig}a shows the total spectrum from ADAFs
around various black hole systems corresponding to 
$m = (10, 10^6, 10^9)$ and $\dot{m} = (10^{-2}, 10^{-4})$.  The
solid, dashed and dotted lines correspond respectively to 
the total emission from the ADAF, the emission from the bulk of
the thermal electrons, and the emission from the $\epm$. As expected,
the low proton temperatures in an ADAF do not allow for a significant
amount of $\epm$ synchrotron emission, and the luminosities are 
too low to be of observational interest.
\subsection{Power--Law Distribution}
For a power--law distribution of protons, 
Figure \ref{fig_all}b shows the total spectrum from ADAFs around the same 
black hole systems corresponding to $m = (10, 10^6, 10^9)$ and
$\dot{m} = (10^{-2}, 10^{-4})$.  
The solid, dashed and dotted lines correspond respectively to 
the total emission from the ADAF, the emission from the bulk of
the thermal electrons, and the emission from the $\epm$, for the
three different values of the proton energy index $s$. 
The heavy solid line corresponds to the total ADAF spectrum 
for $s=2.75$. Unlike the thermal case, the power--law $\epm$ do 
produce significant amounts of detectable emission.

For all black hole masses, Figure \ref{fig_all}b shows 
that bremsstrahlung emission from the thermal electrons 
has a cutoff at $h \nu \sim k T_e$, which
corresponds to $\nu \sim 10^{20}$ Hz, and inverse Compton scattering of the
electrons also turns over at these frequencies due to the decrease in the 
Klein--Nishina cross--section.  The emission for $\nu \gsim 10^{20}$ Hz is  
therefore  only due to proton cooling. 
All spectra show that the $\epm$ naturally
provide a hard power--law tail which extends up to $\sim$ 100 MeV. 
Beyond these energies, the decay of neutral pions into 
$\gamma$--rays dominate the spectrum, and produce a distinct rise in
the spectrum which turns over at a few GeV and extends as a power--law
to very high energies.   An interesting consequence is that although 
the spectral index of the MeV and GeV spectrum are 
different, they are not independent of each other.  For a parent
proton distribution with energy index $s$, the MeV and GeV spectrum 
have spectral dependencies $\nu L_{\nu} \propto \nu^{-s/2}$ and 
$\nu^{-(s-2)}$ respectively (Mahadevan et al. 1997).

For frequencies $\nu \lsim 10^{20}$ Hz, the dominance of the 
proton cooling spectrum, compared with the emission from 
the thermal electrons, depends 
sensitively on the accretion rate 
$\dot{m}$, and to a lesser degree on the mass $m$ of the black hole. 
For low mass black holes (eg. Cyg X--1, A0620) the proton 
cooling spectrum 
is almost always unobservable for all accretion rates.   
For higher mass black holes, however, the proton cooling emission
is observable for $\nu \lsim 10^{11}$ Hz and between 
$10^{14} - 10^{17}$ Hz, and depends 
crucially on $\dot{m}$.

For supermassive black holes, Figure \ref{fig_all}b shows that between frequencies $10^{14} - 10^{17}$ Hz, 
the turn over in the $\epm$ synchrotron spectrum, 
is observable only for low accretion rate systems.  Increasing the
accretion rate, increases the amount of thermal inverse Compton 
scattering ($\propto \dot{m}^3$; Narayan \& Yi 1995b; Mahadevan 1997),
which increases faster than the amount of $\epm$ synchrotron 
radiation ($\propto \dot{m}^2$).   The luminosity
from thermal inverse Compton scattering will therefore dominate the 
spectrum at these frequencies.  The $\epm$ spectrum is therefore 
unobservable for high accretion rate AGN such as NGC 4258 ($\dot{m} \sim 0.01$;
Lasota et al. 1996), while the spectrum dominates the infrared to soft X--ray 
band for low accretion rate systems like our Galactic centre ($\dot{m} \sim 10^{-4}$,
Narayan et al. 1998b; Mahadevan 1998).

For frequencies $\nu \lsim 10^{11}$ Hz, 
the radio spectrum from  ADAFs around
supermassive black holes is substantially modified by the
radiating $\epm$.  Figure \ref{fig_all}b shows that 
the additional synchrotron emission has two
effects: (1) it increases the radio flux by as much as an order of
magnitude at certain frequencies, and (2) changes the shape of the
spectrum by producing an observable spectral break.
As discussed in \S 5,  Mahadevan (1998) has shown that the 
observed spectral break from radio observations of the Galactic
centre (Sgr A$^*$) can be explained quite well by $\epm$ synchrotron radiation.

It is interesting that the synchrotron radiation from the $\epm$
actually produce more emission at low radio frequencies than 
the thermal electrons.  Since the thermal electrons
radiate highly self--absorbed synchrotron emission, the emission at
each frequency in the radio, $\nu_f$, corresponds to black body
emission at a definite radius, $r_f$ (Mahadevan 1997).
Any ``excess'' emission that is greater than the local black body value 
will be highly self--absorbed.  In this case, how does the
inclusion of synchrotron radiation from the $\epm$ increase the radio
emission?  The apparent paradox is resolved by noting that the enhanced
emission at $\nu_f$ results from the high energy $\epm$ radiating
synchrotron emission at larger radii, $r > r_f$.  While any excess
radiation emitted at $r < r_f$ is completely self--absorbed by the
thermal electrons at $r_f$, the $\epm$ at radii $r > r_f$ 
radiate most of
their energy at $\nu \sim \gamma^2 \nu_B \sim \nu_f$, which is now
optically thin to the local black body, and free to escape from the
plasma.  The excess radio emission therefore comes from 
high energy $\epm$ radiating at larger radii.
\section{Discussion and Conclusions} 
ADAFs are the only hot accretion flow models that  attempt to
self--consistently incorporate the viscous hydrodynamics, thermal structure, and  
radiative processes in the inflowing gas.  
The models have been worked out in considerable detail, and the success of 
ADAFs therefore lies in their ability to make robust observational predictions.
The numerous applications of ADAFs to solar and supermassive black holes 
(eg. Di Matteo et al. 1998; Esin et al. 1997, 1998; Fabian \& Rees 1995; Reynolds et al. 1996; Narayan 
et al. 1995; see Narayan et al. 1998a for a review) have not only led us to a 
better  understanding of these systems, but have also reinforced our belief that
ADAFs are a substantial step in the right direction, towards a possibly more 
detailed solution of low luminosity accreting systems. 

In order for the ADAF solutions to exist
two basic assumptions in plasma physics must be satisfied: 
(1) the existence of a hot two temperature plasma, 
and (2) the viscous energy generated primarily heats the protons.
If either of these assumptions fail, then the ADAF solution, and the 
resulting spectra that they produce, cease to exist.  To understand
this, consider first the existence of a two temperature plasma. 
In an ADAF, all the viscous energy is stored in the protons 
of the gas and only a small fraction of this energy is transferred to 
the electrons to be radiated away.  This fraction is determined by 
Coulomb collisions and the large temperature difference of the protons
and electrons.  If some other mechanism (eg. a plasma instability) is 
more efficient in coupling the protons and electrons to equilibrate 
their temperatures, then all the stored energy in the protons, 
would be transferred to the electrons to be radiated away.  No 
energy will be advected with the accreting gas, the 
accretion flow  becomes very luminous, and the ADAF solution 
does not exist.

Second,  the protons become hot and advect the viscous energy 
because viscosity is assumed to primarily heat the protons.
If the viscously dissipated energy went into heating 
the electrons instead,  then this energy would be immediately radiated away, 
resulting in a luminous system.  Again, no energy would be advected by 
the gas, and the ADAF solution does not exist. 
The validity of these assumptions are therefore crucial for the 
existence of ADAFs. 

This paper has focussed on providing additional 
observational tests for these two assumptions, by presenting
detailed spectra from both the protons and electrons.
We have concentrated on the cooling 
of protons via the production of charged pions, which subsequently 
decay into positrons and electrons.  These particles 
interact with the local magnetic fields to produce synchrotron 
radiation.  Until recently, observational signatures of 
hot protons was relegated to very high energies,
where neutral pions decay to produce gamma--rays ($\sim $ few GeV).
The present results not only probe the two temperature 
paradigm further, but also allow us to predict the existence of 
hot protons in a more observationally tractable regime of the 
electromagnetic spectrum. 

It is important to understand the connection between the spectra from the
protons and electrons in ADAFs. Once the spectrum from the thermal electrons
is determined, all the parameters in the ADAF are fixed.  Properties of the
gas such as the numberdensity, temperatures, and magnetic field strengths, cannot be
changed, and the resultant spectra from proton cooling depends
only on the physical processes in the plasma. In particular,
the energies of the created $\epm$, their steady state numberdensity distributions,
and shape of the resulting synchrotron spectra (cf. Fig. \ref{physical_fig}),
depend only on the microphysics of particle collisions, decays,
and radiative processes.   It is this strong interplay between the 
proton and electron spectra that allows testable predictions  of the 
basic assumptions of ADAFs,  and also allows a probe into the proton  energy 
distribution in these flows.

Since it is not understood whether viscous heating produces 
a thermal or power--law distribution of proton energies, we 
have calculated the $\epm$ spectrum due to both distributions;
the true distribution will be some combination of 
the two, and can be  determined by taking the weighted sum of 
the thermal and power--law spectra.
Recently, there have been a number of theoretical
investigations which consider particle heating in ADAFs
(Bisnovatyi--Kogan \& Lovelace 1997; Blackman 1998; Quataert 1998;
Gruzinov 1998; Quataert \& Gruzinov 1998; see Narayan et al. 1998a for
a review), some of which also argue for and against both proton distributions. 
Here, we have attempted to answer this question observationally 
by providing accurate spectra from these distributions.

If the proton distribution is thermal, the resulting $\epm$ synchrotron 
spectrum peaks between frequencies $10^{12}-10^{17}$ Hz, and produces very 
low luminosities.  In addition, the radiation
by the thermal electrons dominates the emission from the
radio to hard X--rays, which  results in an $\epm$ spectrum that is completely 
unobservable (Fig. \ref{thermalfig}a).  In this case, the only observable 
signature of hot protons would be 
in the gamma--rays, but the predicted fluxes are much too 
low to be detected even by the next generation gamma--ray telescope, GLAST.

However, if the protons have a power--law distribution of energies, 
the $\epm$ synchrotron spectrum can be observed. 
For supermassive black holes, the emission appears in the radio,
and, depending on the accretion rate, also between infrared and soft X--ray 
energies.  In particular, the best systems to look 
for such radiation would be in supermassive black holes 
with low accretion rates.  Fortunately, the supermassive black 
hole at our Galactic centre (Sgr A$^*$) offers such an opportunity.  
It is the closest low luminosity AGN that we know, and 
has been observed quite extensively.

Recently, Narayan et al. (1998b) have used an ADAF model 
to explain the spectrum and low luminosity from \sgr.  
For the canonical values of $\alpha=0.3$ and $\beta=0.5$, and the
dynamically measured mass $M=2.5\times 10^6 \, M_{\odot}$ (Haller et al. 1996; 
Eckart \& Genzel 1997), they vary just one free parameter, $\dot{m}$, to 
fit the X--ray flux, and 
show that the resulting spectrum from the thermal 
electrons explains the
low luminosity and observed broad band spectrum quite well.  The model, however, 
has difficulties in explaining the non--uniform radio spectrum. 
It is important to realize that once the thermal electron spectrum is fixed,
all the parameters in an ADAF are determined, and there are no
more free parameters available.  The non--uniform radio spectrum has 
therefore always been unexplained by the ADAF model.

This discrepancy has recently been solved by considering $\epm$ synchrotron 
emission from the ADAF around \sgr (Mahadevan 1998). Without changing 
any parameters, and just including another physical process -- $\epm$ synchrotron emission from 
a power--law proton distribution,
Mahadevan (1998) shows that the ADAF model naturally reproduces the 
observed radio spectral break at $\sim 86$ GHz, and
accounts for the low frequency luminosities.  This is shown by the solid line in
Figure \ref{sgr},
which provides quite compelling evidence that a two temperature 
plasma probably does exist in an ADAF around  Sgr A$^*$.  
More importantly, the results provide, for the first time,  
observational support for the basic assumptions in an ADAF 
(Mahadevan 1998).

While one of the ADAF assumptions requires almost all of the viscous energy to 
be transferred to the protons, it does not indicate whether this energy
should heat the protons into a power--law or thermal distribution.  This depends on 
the details of viscous heating, and it is interesting that the recent results from \sgr 
allows an observational answer to this question.
If we characterize the amount of viscous energy that goes into  a 
power--law by $\Delta$, then the case when no energy goes into
the power--law distribution is $\Delta = 0$, which is shown in Figure
\ref{thermalfig}a; $\Delta = 1$ corresponds to Figure \ref{fig_all}b.  
The question arises as to what the most likely values of 
$\Delta$ can be, ie. how much of the viscous heating is 
transferred to a power--law distribution and how much to 
a thermal one.  To assess this, we use the results of Mahadevan (1998)
as our baseline model. $\Delta = 1$ corresponds to the solid line in Figure \ref{sgr},
while $\Delta = 0$ corresponds to the long dashed line (Narayan et al. 1998b). 
The dotted, short dashed, and dot dashed lines correspond to spectra for $\Delta = 0.5, 0.25,$ and 
0.1 respectively.    

Figure \ref{sgr} shows that the results are not very sensitive to $\Delta$;  this 
factor will only appear linearly in equation (\ref{zetaeq}).  
While the luminosity of the optically thin
part of the power--law spectrum decreases linearly with $\Delta$,
the self--absorbed radio part is less  sensitive.  
For $\Delta \gsim 0.3$, the agreement with the radio spectrum does
not change significantly.  This implies that
the observations are consistent with approximately a third or more of the
viscous energy going into a power--law distribution of protons, and the rest
into a thermal or quasi--thermal one.  The results therefore provide clues to the physics 
of viscous heating in these flows, and could be used as
tools to aid future theoretical work in resolving some of the complex questions in
plasma physics.

The results of Mahadevan (1998) encourages us to adopt the
view that the protons in an ADAF, might in fact 
have a power--law distribution in energies.
This gives rise to numerous testable predictions in the fluxes 
in various wavebands as well as the spectral shapes.  
First, all spectra from two body emission processes, in the optically thin limit,
must have constant ratios in their fluxes.
This is a result of all two body processes varying as the square of the 
density.  The three processes in an 
ADAF, which depend on the square of the density, 
are bremsstrahlung emission and
the creation of neutral and charged pions.  From Figure 
\ref{fig_all}b it is clear 
that these three processes always have the same relative amplitudes in
the optically thin regime ($\nu \gsim 10^{10}$ Hz). In particular the 
ratio of the gamma--ray to bremsstrahlung to $\epm$ synchrotron fluxes 
are all the same, and are determined essentially by the ratios of their respective
cross--sections and rates.  While the cross--section for
charged pion creation is greater than that for neutral pions, the
synchrotron spectrum from the $\epm$ has a lower peak flux than the
$\gamma$--rays.  This is due to the synchrotron radiation being
emitted over a much broader frequency range. 

For low accretion rate systems, these emission processes dominate the radiation for 
frequencies $\nu \gsim 10^{12}$ Hz, which provides 
strong testable correlations among
the fluxes in the optical, X--rays, hard X--rays, and $\gamma$--ray 
bands. Further, changes in the accretion rate, leads to changes in the
the densities, which would result in simultaneous variations in the fluxes at these different 
frequencies.  Observations of such variability would provide very interesting tests
of the ADAF paradigm. It is apparent from Figure \ref{fig_all}b that the best laboratories to
verify these predictions would be in low luminosity AGN.

In addition to the predicted correlation in fluxes, there are also
correlations in the expected spectral shapes and slopes. As discussed
in \S 4,  the ubiquitous presence of a hard power--law at $\sim $MeV
and $\sim $ GeV energies provides two independent checks on the
distribution of protons. 
In an ADAF the detection of one requires the presence of the other.
The power--law spectra presented here do not show any high energy cutoff.
This is a direct consequence of postulating a
power--law proton distribution which extends to very high 
energies.  However, if we arbitrarily postulate a maximum proton energy 
(for example the energy at which the
proton gyroradius equals the size of the ADAF) then the resulting
spectra would also show a corresponding cutoff.

The increased radio flux, from supermassive black holes, 
 due to $\epm$ radiation,  
facilitates easier comparisons with observations, and can be used to constrain 
the sizes of ADAFs in AGN.  
Recently, Herrnstein et al. (1998) have compared observational upper
limits to the 22 GHz emission from NGC 4258 with an ADAF model, and
conclude that an ADAF cannot have radii larger than $\sim 100$
Schwarzschild. We might expect that the current results would produce
more emission in the radio, and the size of the ADAF would therefore be
further reduced. However, this is not the case since in order to
produce more emission at 22 GHz, requires synchrotron radiation from
$\epm$ at radii $r \gsim 100$.   In fact the current results allows for 
a more definite upper limit to the size of the ADAF deduced by Herrnstein et al. 
(1998), since a larger outer radius implies more emission at 22 GHz from the
$\epm$.   This technique can also be used to constrain, or establish, the 
existence of ADAFs at the cores of nearby low luminosity ellipticals.

Finally, we discuss two physical processes that have been neglected: (1) 
inverse Compton scattering of soft photons by the created $\epm$, and 
(2) the observational importance of an annihilation line.
First, the significance of inverse Compton scattering, by the created $\epm$, 
is determined by comparing the ratio of the total 
inverse Compton power to synchrotron power, $P_{\rm Com}/P_{\rm syn} = U_{\rm ph}/U_{B}$, where
$U_{\rm ph}, \, U_{B}$ are the photon and magnetic field energy densities respectively.
For an ADAF, $U_{\rm ph}/ U_{B} \sim L_{\rm emitted}/L_{\rm advected} \ll 1$, where 
$L_{\rm emitted}, \, L_{\rm advected}$ are the emitted and advected luminosities, respectively, and
inverse Compton scattering of the $\epm$ is therefore not energetically important.

Second, assessing the observational importance of an annihilation line requires a knowledge of 
the total energy in the line, as well as the line width.  Since the annihilation cross--section
is greatest for Lorentz factors $\gamma \sim 1$, and most of the synchrotron emission occurs from 
electrons with $\gamma \sim 200$, the power in the annihilation line will be $\sim P_{\rm syn}/200$.  In the
case of \sgr  this corresponds to $\sim 10^{33}$ erg s$^{-1}$.  If the width of the line is 
narrow, then the line will be observable.  However, the $\epm$ that annihilate have
thermalized with the ambient electrons at temperatures $\sim 10^{9.5}$K.    At such 
high temperatures, the annihilation line is sufficiently broadened ($\Delta E/E =  \sqrt{2kT_e/m_ec^2} 
\sim 1$ MeV), which results in a spectral line that is unobservable.

\vspace{.3in}
\noindent {\em Acknowledgments}:  I thank Ramesh Narayan and Andy Fabian for useful discussions.
\newpage

\newpage
\setcounter{figure}{0} 
\newpage
\begin{figure}
\vspace{6.5in}
\vspace{1.in}
\centerline{
\includegraphics{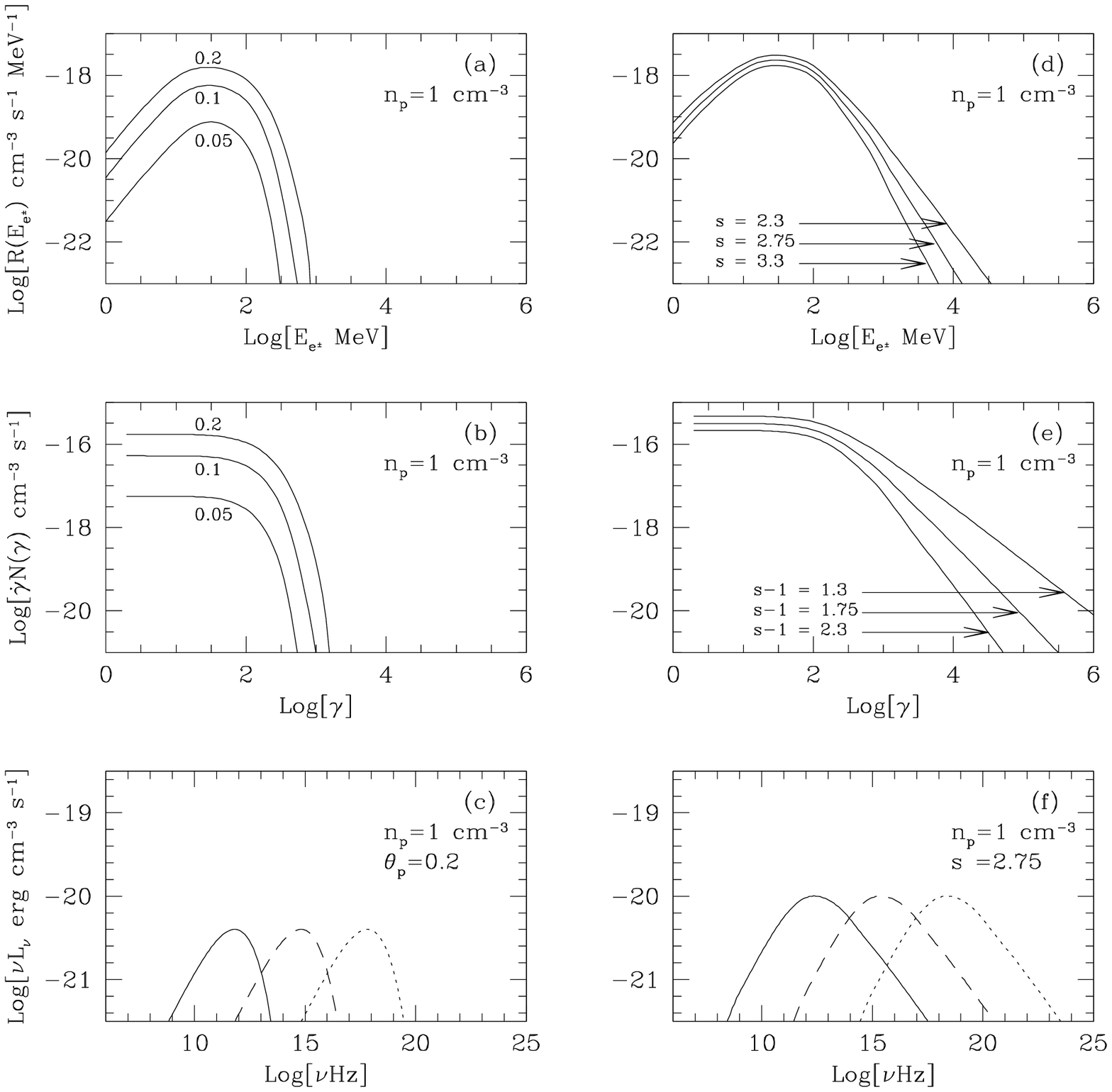}
}
\vspace{-.25in}
\caption[Physical_Fig]{
Figures showing the microphysics of $\epm$ production, and the resulting synchrotron radiation. 
The left column is for a thermal distribution of protons where the curves are labeled
by the dimensionless proton temperature, $\theta_p = kT_p/m_pc^2$.  The right column is for a power--law distribution 
for three values of the energy index $s$.  
Figures (a) and (d) correspond the rate of creation of $\epm$ for density of protons equal
to unity.  For a numberdensity $n$, the vertical axis must be multiplied by $n^2$.  Figures 
(b) and (e) show the resulting steady state distribution $N(\gamma)$ of $\epm$.  The y--axis 
plots $C(\gamma) = \dot{\gamma} N(\gamma)$ (cf. eq. 10).  The synchrotron 
emission from the steady state distribution is shown in figures (c) and (f; cf. eq. 11).
}
\label{physical_fig}
\vspace{-.5in}
\end{figure}

\newpage
\begin{figure}
\vspace{6.5in}
\centerline{
\includegraphics{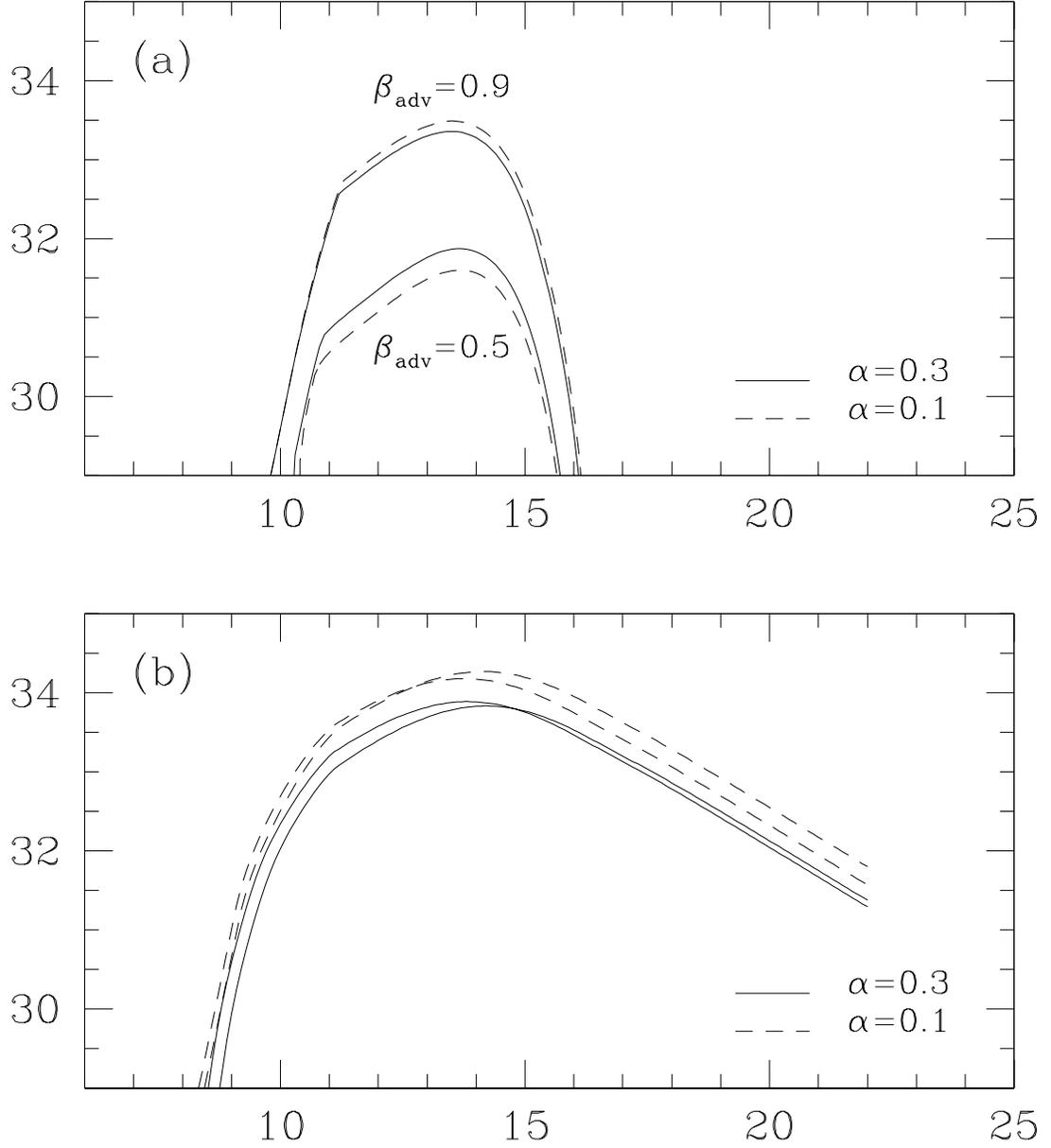}
}
\vspace{-.25in}
\caption[Changes]{
Positron and electron synchrotron spectrum from an ADAF with $m= 10^6$, $\dot{m} = 10^{-4}$, for (a) a 
thermal distribution of protons, and (b) a power--law distribution with energy index $s=2.75$.  
In the case of a power--law distribution,
the spectrum which turns over at low (high) frequencies correspond to $\beta_{\rm adv} = 0.5 \ (0.9)$.
}
\label{fig2}
\vspace{-.5in}
\end{figure}

\newpage
\begin{figure}
\vspace{6.5in}
\centerline{
\includegraphics{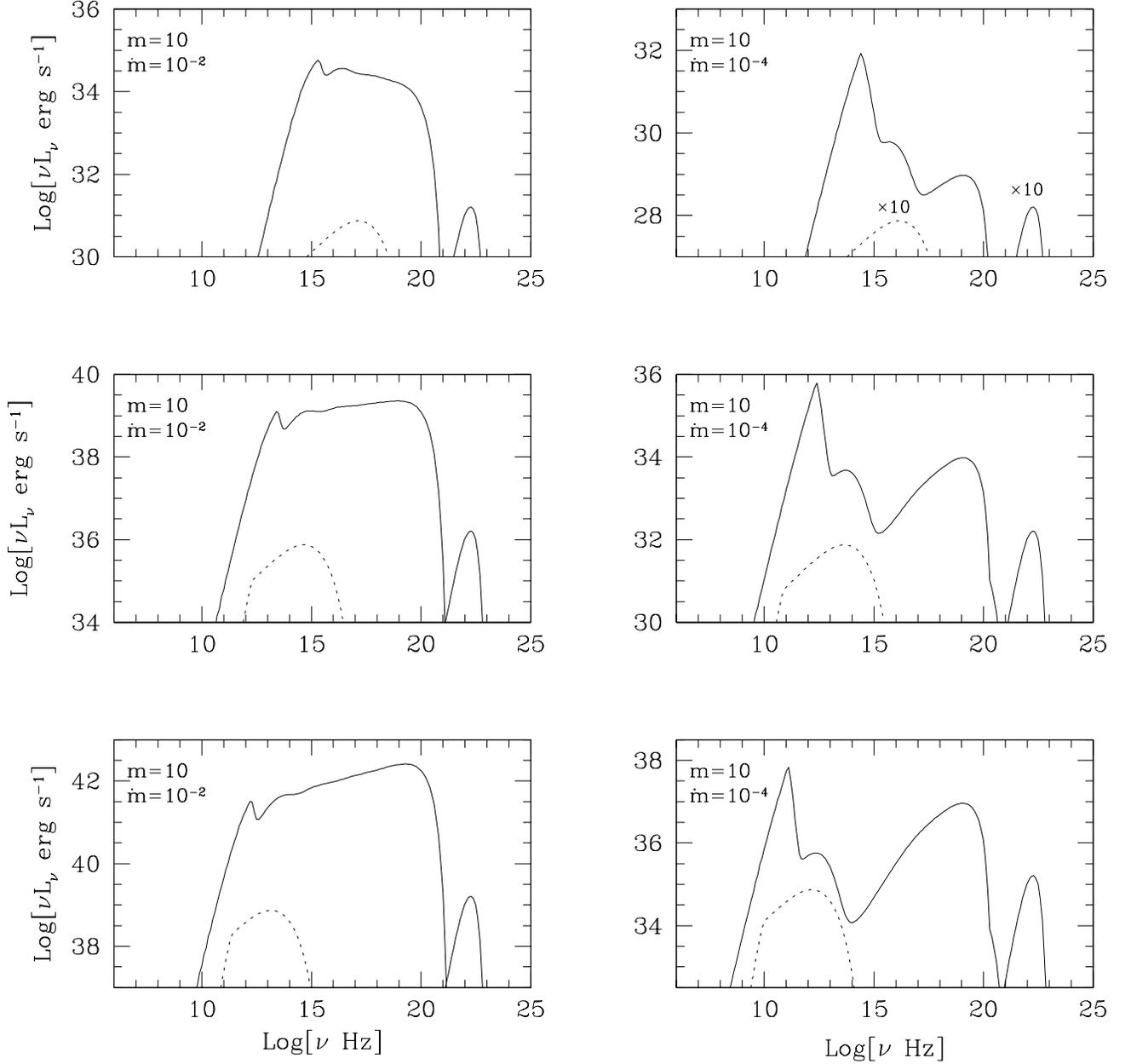}
}
\vspace{-.25in}
\caption[Thermal]{(a)
Complete ADAF spectra (solid line) using a thermal distribution of protons and electrons, 
for different black hole masses and accretion rates.  
In all the calculations $\alpha = 0.3$, $\beta_{\rm adv} = 0.5$. 
The $\epm$ synchrotron spectrum (dotted line) is unobservable in all cases.  
For $(m,\dot{m}) = (10, 10^{-4})$, the
$\gamma$--ray spectrum and $\epm$ synchrotron spectrum have been multiplied by 10; the 
actual fluxes are ten times lower than shown.
}
\label{thermalfig}
\vspace{-.5in}
\end{figure}

\addtocounter{figure}{-1}

\newpage
\begin{figure}
\vspace{6.5in}
\centerline{
\includegraphics{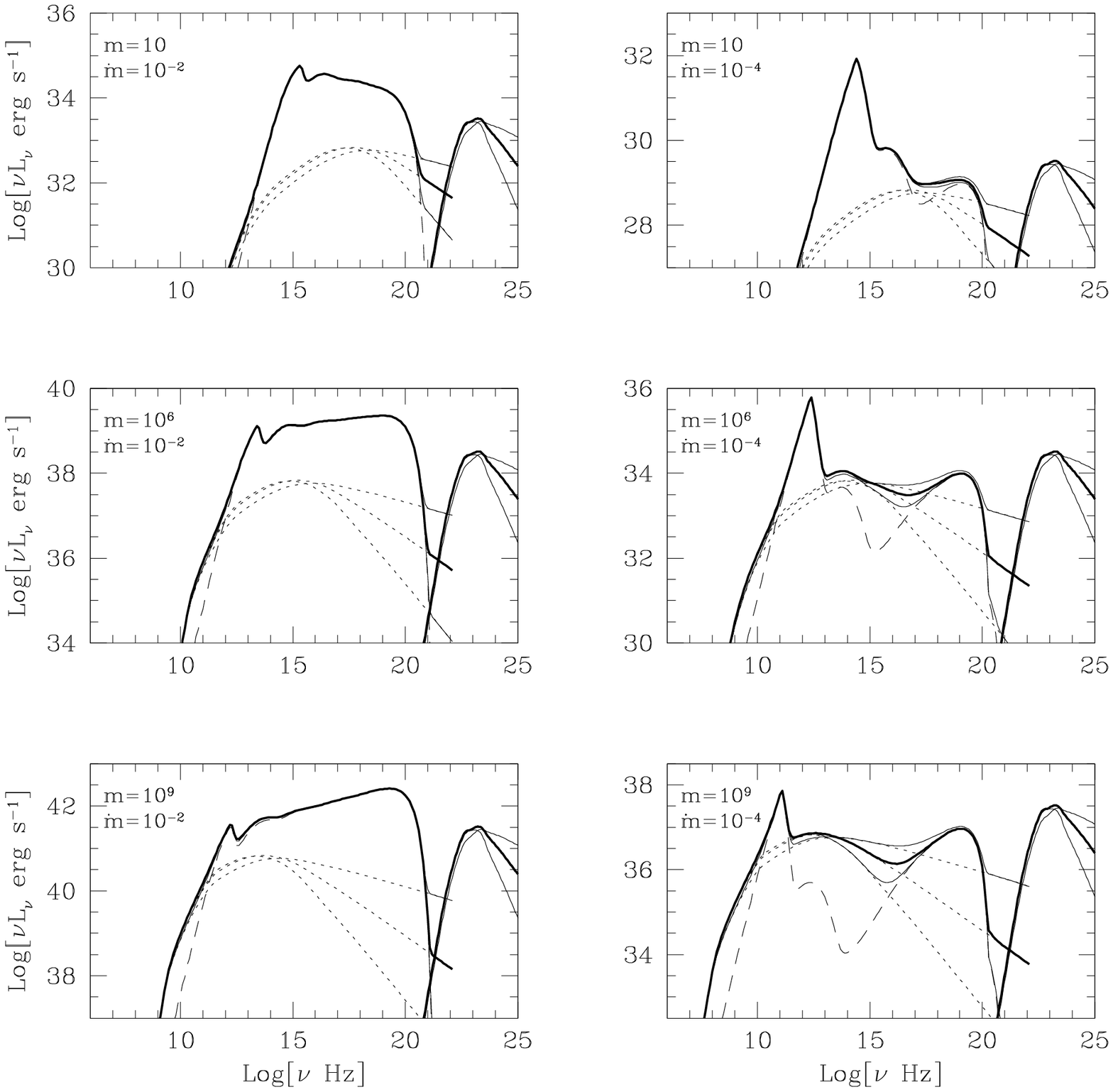}
}
\vspace{-.25in}
\caption[]{(b)
Complete ADAF spectra (solid lines) using a power--law distribution of protons and thermal electrons, 
for the same parameters given in figure 3a.  Three spectra corresponding to 
proton energy indices $s=2.3, \ 2.75, $ and 3.3 are shown.  The heavy solid line represents
$s= 2.75$, with the spectra below and above corresponding to $s=3.3$ and $s=2.3$ respectively.
The dotted and dashed lines correspond  to the emission from the $\epm$ and the thermal electrons
respectively.  
}
\label{fig_all}
\vspace{-.5in}
\end{figure}

\newpage
\begin{figure}
\vspace{6.5in}
\centerline{
\includegraphics{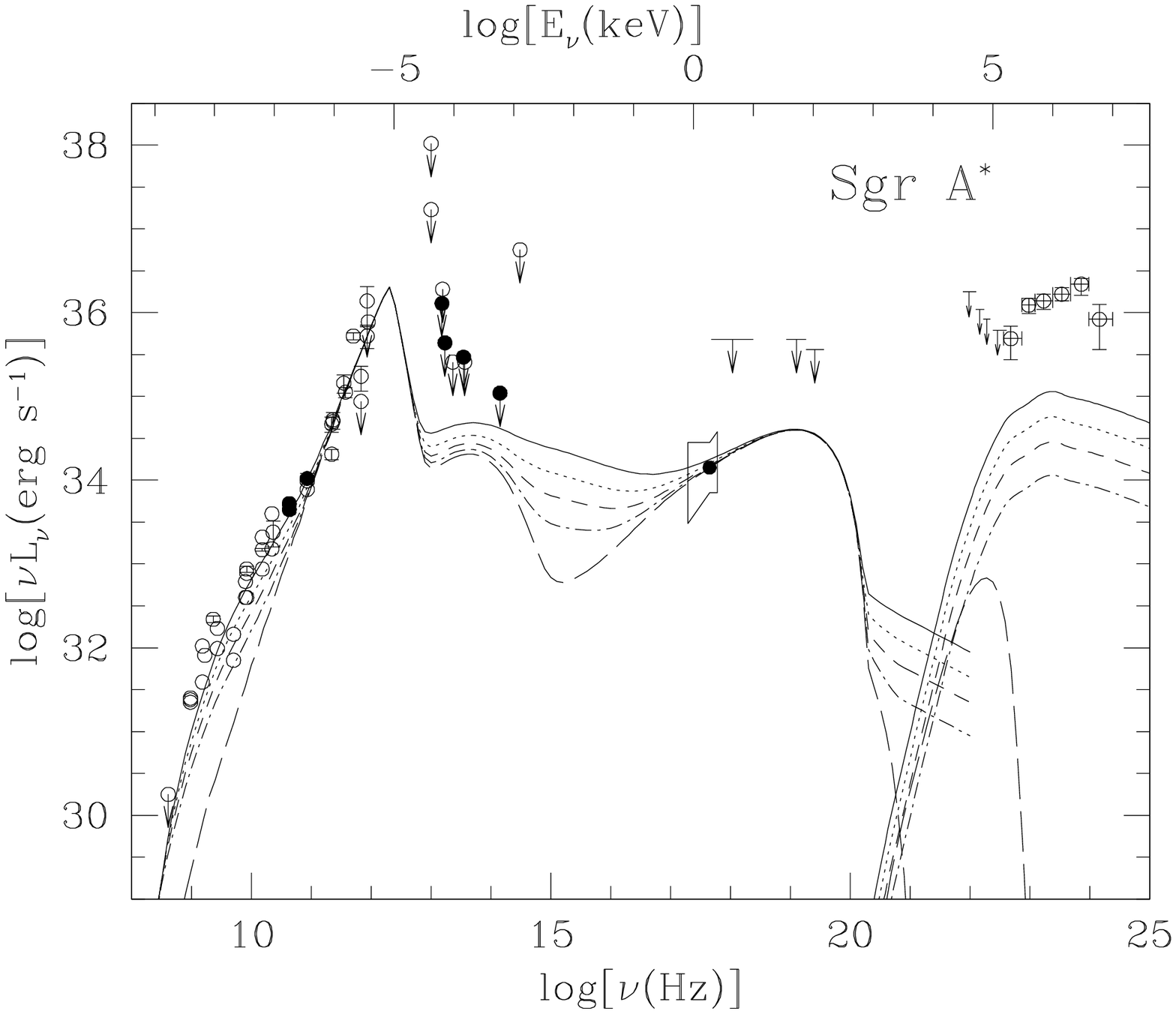}
}
\vspace{-.25in}
\caption[]{
The spectrum from an ADAF around Sgr A$^*$ for different values of the parameter $\Delta$.  The 
data are taken from Narayan et al. (1998b).
$\Delta$ is the fraction of viscous energy that heats the protons into a power--law distribution. 
The solid, dotted, small dashed, dot--dashed, and long dashed lines (from high to low luminosity)
corresponds to $\Delta = 1.0, 0.5, 0.25, 0.1,$ and, 0.0 respectively. The spectrum corresponding
to $\Delta = 1$ (solid line) is taken from Mahadevan (1998).
}
\label{sgr}
\vspace{-.5in}
\end{figure}
\end{document}